\documentclass[twocolumn,showpacs,prl,superscriptaddress]{revtex4}

\usepackage{graphicx}

\renewcommand{\epsilon}{\varepsilon}
\newcommand{\figurewidth}{0.45\textwidth}
\newcommand{\narrowfigurewidth}{0.22\textwidth}

\begin{document}
\title{Simulation of phase transitions in highly asymmetric fluid mixtures}
\author{Jiwen Liu}
\affiliation{Dept.\ of Materials Science and Engineering,
University of Illinois at Urbana-Champaign, Urbana, Illinois 61801, U.S.A.}
\author{Nigel B. Wilding}
\affiliation{Department of Physics, University of Bath, Bath BA2 4LP, U.K.}
\author{Erik Luijten}
\affiliation{Dept.\ of Materials Science and Engineering,
University of Illinois at Urbana-Champaign, Urbana, Illinois 61801, U.S.A.}

\date{\today}

\begin{abstract}
  
  We present a novel method for the accurate numerical determination of
  the phase behavior of fluid mixtures having large particle size
  asymmetries. By incorporating the recently developed geometric cluster
  algorithm within a restricted Gibbs ensemble, we are able to probe
  directly the density and concentration fluctuations that drive phase
  transitions, but that are inaccessible to conventional simulation
  algorithms. We develop a finite-size scaling theory that relates these
  density fluctuations to those of the grand-canonical ensemble, thereby
  enabling accurate location of critical points and coexistence curves
  of multicomponent fluids. Several illustrative examples are presented.

\end{abstract}

\pacs{05.10.Ln, 64.70.Fx, 64.70.Ja, 82.70.Dd}

\maketitle

The vast majority of commercially relevant fluids are multicomponent
mixtures. An understanding of the phase behavior of these systems is of
paramount importance for applications, and also a matter of great
fundamental interest~\cite{larson99}.  With the advent of powerful
computers, various computational techniques have been devised to
directly determine fluid phase behavior~\cite{frenkel-smit2}. However,
these methods are all restricted to fluids in which the various
components have similar sizes, whereas important phenomena occur in
highly size-asymmetric multicomponent fluids such as colloidal
dispersions, colloid-nanoparticle mixtures, and polymer
solutions~\cite{larson99,tohver01}.

The computational bottleneck for existing simulation methods arises from
the fact that they cannot simultaneously relax a fluid system on
disparate length scales. Specifically, for large size ratios, the big
particles become `jammed' by the smaller ones. Recently, however,
building on earlier work~\cite{dress95}, a geometric cluster Monte Carlo
algorithm (GCA) was proposed~\cite{geomc,liu05a} that facilitates
\emph{rejection-free} simulations of highly size-asymmetric mixtures via
large-scale collective updates which move whole groups of particles in a
single step.  Although this method has been successfully applied to
problems relating to colloidal stabilization~\cite{liu04b,martinez05},
in which size asymmetries span several orders of magnitude, it is
incapable of dealing with the density and concentration fluctuations
associated with phase separation, since it inherently operates in a
canonical ensemble in which the particle number and volume are fixed.
This limitation cannot be overcome by incorporating cluster moves into a
standard grand-canonical~(GC) or constant-$NpT$ ensemble, because this
does not address the underlying shortcomings of these ensembles with
respect to sampling of density fluctuations in asymmetric mixtures.

It is the purpose of this Letter to introduce a method that overcomes
these problems. This is achieved by embedding cluster moves in a variant of
the Gibbs ensemble~\cite{panagiotopoulos87,mon92}, in such as way that
they \emph{couple} to the density fluctuations, resulting in efficient
exploration of configuration space.  To exploit this approach we present
a finite-size scaling theory that permits the determination of the
critical point and the phase boundary. As an illustration, we apply the
method to study liquid-vapor coexistence in asymmetric binary mixtures,
for which we show that the presence of even small quantities of
small-particle additives can strongly affect the location of the
critical point. Furthermore, depending on the nature of the interaction
of the additive with the fluid particles, the critical temperature can
either be enhanced or depressed.

To enable density fluctuations, we distribute a prescribed number of
particles~$N_0$ over two boxes, and devise an operation that exchanges
particles between these boxes~\cite{panagiotopoulos87} to maintain
chemical equilibrium.  By adopting the symmetrical \emph{restricted}
Gibbs (RG) ensemble~\cite{mon92}, in which the boxes have equal
\emph{constant} volumes~$V=L^d$, geometric cluster moves can be used for
this exchange operation. The prescription for a cluster move closely
follows the original algorithm~\cite{geomc}, with the crucial difference
that a geometric operation not only alters the position of a particle,
but also transfers it from one box to the other.  Specifically, a pivot
is placed at a random position within the first box, as well as at the
corresponding position within the second box. A particle~$i$ is picked
at random from one of the boxes (denoted 1) and point-reflected with
respect to the pivot from its original position~$\mathbf{r}_i$ to
$\mathbf{r}_i'$. However, instead of placing the particle
at~$\mathbf{r}_i'$, we place it at the corresponding point
$\mathbf{\bar{r}}_i'$ in the other box~(denoted 2), subject to periodic
boundary conditions.  Thereafter, any particle~$j$ interacting with
particle~$i$ around its original position in box~1 or its new position
in box~2 will also be considered for point reflection around the pivot
and subsequent transfer to the opposite box, with a probability $p_{ij}
= \max[1 - \exp(-\beta \Delta_{ij}), 0]$, where $\Delta_{ij} = -
V(|\mathbf{r}_i-\mathbf{r}_j|)$ if particles $i$ and~$j$ originally
reside in the same box and $\Delta_{ij} =
V(|\mathbf{\bar{r}}_i'-\mathbf{r}_j|)$ if $i$ and~$j$ originally reside
in different boxes. $V(r)$ denotes a general pair potential and $\beta$
the inverse temperature $1/k_{\rm B}T$.  Note that~$p_{ij}$
\emph{solely} depends on the pair interaction between particles $i$
and~$j$, rather than on the total energy change resulting from the
displacement of particle~$j$. The cluster construction proceeds
iteratively for all particles interacting with each particle~$j$.  Upon
completion of a cluster move, a new pivot is selected.  The proof of
detailed balance is analogous to that for the generalized
GCA~\cite{geomc,liu05a}.

The exchange of particles between boxes leads to (complementary) density
fluctuations around the average number density $\rho_0=N_0/(2V)$ in each
box. The fluctuation spectrum of the number density in box~$1$,
$\rho_1=N_1/V$, can be related to the \emph{grand-canonical} probability
distributions~$P$ of the number densities of both boxes
via~\cite{bruce97}
\begin{equation}
  P^{RG}(\rho_1|\rho_0,V,T) \propto
  P(\rho_1|\mu,V,T)P(2\rho_0-\rho_1|\mu,V,T) \;,
  \label{eq:map}
\end{equation}
where we note that $P^{RG}(\rho_1)$ is
independent of the choice of chemical potential $\mu$ 
on the right-hand side~\cite{bruce97}.

Since the right-hand side of Eq.~(\ref{eq:map}) is the product of a
function and a shifted (by $2\rho_0$) and reflected form of this
function, $P^{RG}(\rho_1)$ is symmetric (even) with respect to its mean
$\bar{\rho}_1=\rho_0$. To facilitate comparison of the forms of
$P^{RG}(\rho_1)$ for various choices of $\rho_0$, it is expedient to
consider distributions of zero mean, to which end we define
$x=\rho_1-\rho_0$ and write
\begin{equation}
  P^{RG}(x|\rho_0,T) \propto P(x+\rho_0|T)P(-x+\rho_0|T) \;,
  \label{eq:RGSHIFT}
\end{equation}
where we have suppressed reference to the constant volume $V$ and the
(arbitrary) chemical potential $\mu$.

The parameter space of the RG ensemble is spanned by $\rho_0$ and $T$.
In the vicinity of the critical point ($\rho_0^c, T^c$) terminating a
line of liquid-vapor coexistence of a pure fluid or fluid mixture,
$P^{RG}$ exhibits universal scaling behavior. Introducing reduced
variables $\varrho_0 \equiv (\rho_0-\rho_0^c)/\rho_0^c$ and $t \equiv
(T-T^c)/T^c$, we make the following \emph{ansatz} for the finite-size
scaling (FSS) properties of $P^{RG}(x)$,
\begin{equation}
  P_L^{RG}(x|\varrho_0,t) = a_0L^{\beta/\nu}
  \tilde{P}^{RG}(a_0L^{\beta/\nu}x, a_1\varrho_0L^{\beta/\nu},a_2tL^{1/\nu}) \;.
  \label{eq:ansatz}
\end{equation}
Here $\tilde{P}^{RG}$ is a universal function which is \emph{symmetric}
in $x$ for all values of $t$ and $\varrho_0$; $\beta$ and $\nu$ are
critical exponents, and $a_0,a_1,a_2$ are nonuniversal metric factors.
The arguments in $t$ and $\varrho_0$ control deviations from criticality.
The temperature field has the form familiar from the FSS
properties of magnets~\cite{binder81} or fluids~\cite{bruce92}, while
that in $\varrho_0$ is particular to the RG ensemble. As
one can verify~\cite{note-longpaper} from an expansion of
Eq.~(\ref{eq:RGSHIFT}) with respect to $\rho_0$, together with the
known~\cite{bruce92} symmetry properties of the derivatives of $P(x)$,
variations in the form of $P^{RG}(x)$ are---to leading
order---controlled by the value of $\rho_0^2$; all terms having odd
powers of $\rho_0$ are antisymmetric in $x$ and hence absent on symmetry
grounds.

To characterize the form of $\tilde{P}^{RG}(x)$ as a function of
$\varrho_0$ and $t$, it is useful to consider the behavior of the
dimensionless fourth-order cumulant ratio $Q \equiv \langle
x^2\rangle^2/\langle x^4\rangle$~\cite{binder81}, whose scaling
properties follow from Eq.~(\ref{eq:ansatz}) as
\begin{equation}
  Q_L(\varrho_0,t)=\tilde{Q}(q_1\varrho_0L^{\beta/\nu},q_2tL^{1/\nu})\:,
  \label{eq:Qscale}
\end{equation}
with $\tilde{Q}$ a universal function and $\tilde{Q}(0,0)\equiv Q^*$.
The value of $Q^*=0.711901$ is known \emph{a priori} by virtue of the
result that $\tilde{P}^{RG}(a_0L^{\beta/\nu}x,0,0) \propto
\left[P^*(L^{\beta/\nu}m)\right]^2$~\cite{bruce97}, with
$P^*(L^{\beta/\nu}m)$ the universal critical 
Ising magnetization distribution~\cite{tsypin00}.  Measurements of
$Q_L(\varrho_0,t)$ for a range of global densities~$\rho_0$ provide a
useful route to locating criticality.  Specifically, consider the locus
of points in $\varrho_0$--$t$ space for which $Q_L(\varrho_0,t)=Q^*$,
which we term the ``iso-$Q^*$ curve.'' Expanding Eq.~(\ref{eq:Qscale})
with respect to $t<0$ and $\varrho_0$, and recalling that only terms
involving even powers of $\varrho_0$ are nonzero, one has
\begin{equation}
  Q_L(\varrho_0,t) = Q^* 
  \left[
    1 +q_1 \varrho_0^2L^{2\beta/\nu} +q_2 tL^{1/\nu}+
    \mathcal{O}(\varrho_0^4,t^2)
  \right] \:,
  \label{eq:Qexp}
\end{equation}
from which it follows that, sufficiently close to the critical point,
the iso-$Q^*$ curve is a \emph{parabola} in $\varrho_0$--$t$
space,
\begin{equation}
  \varrho_0^2 = -(q_2/q_1)L^{(1-2\beta)/\nu}t \;.
  \label{eq:isoqscale}
\end{equation}
The maximum of this parabola (at $\varrho_0 = t = 0$) coincides with the
critical point and hence, by fitting to a few estimated points on the
iso-$Q^*$ curve, one can readily determine the critical parameters 
$\rho_0^c$ and $T^c$.

Turning now to the task of obtaining subcritical coexistence
properties within the RG framework, we consider the peak
positions of $P^{RG}(\rho_1|\rho_0,t)$ on some (subcritical) isotherm. One
finds that when $\rho_0$ equals the coexistence diameter $\rho_d\equiv
(\rho_g+\rho_l)/2$, with $\rho_g$ and $\rho_l$ the gas and liquid
densities, the peak positions of $P^{RG}(\rho_1)$ coincide with the
coexisting densities, which can thus be simply read off from a measured
histogram of its form~\cite{note-diameter}. An effective method for
locating the diameter~$\rho_d$ exploits the fact that the even moments
of $P^{RG}(x|\rho_0,t)$ are maximized when $\rho_0=\rho_d$. From the
absence of odd powers of $\rho_0$ in the expansion of
Eq.~(\ref{eq:RGSHIFT}), it can then be shown~\cite{note-longpaper} that
the variance $\sigma^2(\rho_0|t)$ of $P^{RG}(x|\rho_0,t)$ varies to
leading order quadratically in $(\rho_0-\rho_d)$, i.e.,
\begin{equation}
  \sigma^2(\rho_0|t)= \sigma^2(\rho_d,t)-b(\rho_0-\rho_d)^2 \;,
  \label{eq:m2-parabola}
\end{equation}
with $b$ a positive constant. By fitting to estimates of
$\sigma^2(\rho_0|t)$, this result facilitates a determination of $\rho_d$
and thence the coexisting densities.

\begin{figure}
  \includegraphics*[width=\figurewidth]{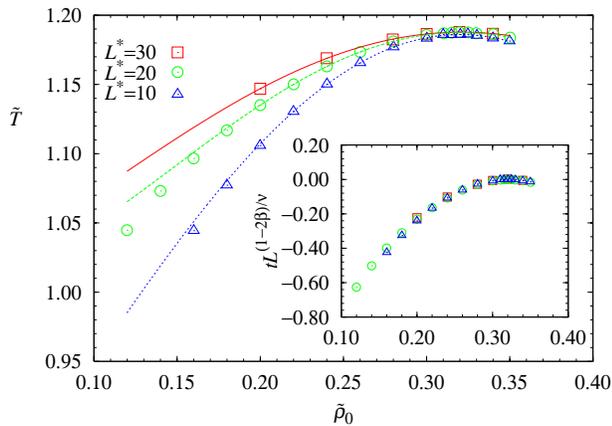}
  \caption{(color online) Measured iso-$Q^*$ points (symbols) for a pure
  LJ fluid, together with parabolic fits (curves). The maxima
  of the curves (shown for $L/\sigma = 10$, $20$, $30$) in the
  $\tilde{\rho}_0$--$\tilde{T}$ plane correspond to the critical point.
  The data collapse in the inset confirms the scaling prediction
  [Eq.~(\ref{eq:isoqscale})].}
  \label{fig:iso-q}
\end{figure}

To test the scaling theory, we first perform simulations using the GCA
in the restricted Gibbs ensemble for a pure Lennard-Jones (LJ) fluid. We
employ a potential cutoff $2.5\sigma$ and reduced system sizes $L^* =
L/\sigma = 10$, $20$, $30$.  At fixed global density~$\rho_0$, histogram
reweighting can be used to determine a point on the iso-$Q^*$ curve,
i.e., the temperature at which $Q_L(\rho_0,T)$ takes the value~$Q^*$.
Repeating for a range of $\rho_0$ values allows the entire iso-$Q^*$
line to be mapped.  As shown in Fig.~\ref{fig:iso-q}, the iso-$Q^*$
curves for the various $L^*$ are indeed parabolas [cf.\ 
Eq.~(\ref{eq:isoqscale})] that coincide at their maximum.  A careful
analysis~\cite{note-longpaper} of the position of this maximum, $\tilde
T^c =k_{\rm B}T^c/\varepsilon = 1.1878~(2)$, $\tilde \rho^c = \rho^c_0
\sigma^3 = 0.3204~(5)$, reveals excellent agreement with an existing GC
estimate, $\tilde T^c = 1.1876~(3)$, $\tilde\rho^c =
0.3197~(4)$~\cite{wilding95}.  The inset confirms the finite-size
scaling predicted in Eq.~(\ref{eq:isoqscale}) with remarkable accuracy.
Furthermore, the critical density distribution~$\tilde{P}^{RG}(x)$ is indeed in
quantitative agreement (not shown) with the square of the critical Ising
magnetization distribution, as predicted~\cite{bruce97}.

Having established the validity of our methodology, we now exploit the
strengths of the GCA to address a typical problem that is intractable
for conventional simulations in the GC, $NpT$, or Gibbs ensemble. We
consider a binary, strongly size-asymmetric mixture of LJ particles of
size~$\sigma$ and small particles (``additives'') of diameter $\sigma_s
= \sigma/10$. Depending on their interaction with the large particles,
the presence of additives will affect the phase behavior and shift the
location of the liquid-vapor critical point compared to the pure fluid.
The additives mutually interact via a weakened LJ potential,
\begin{equation}
  V_{ss}(r) = 4 \left(\frac{\varepsilon}{10}\right)
              \left[ \left(\frac{\sigma_s}{r}\right)^{12} -
                     \left(\frac{\sigma_s}{r}\right)^6 \right] 
  \quad (r < 2.5\sigma_s) \;,
  \label{eq:lj-small}
\end{equation}
whereas a large and a small particle interact as hard spheres at a
separation $\sigma_{ls} = (\sigma + \sigma_s)/2$.

\begin{figure}
  \includegraphics*[width=\figurewidth]{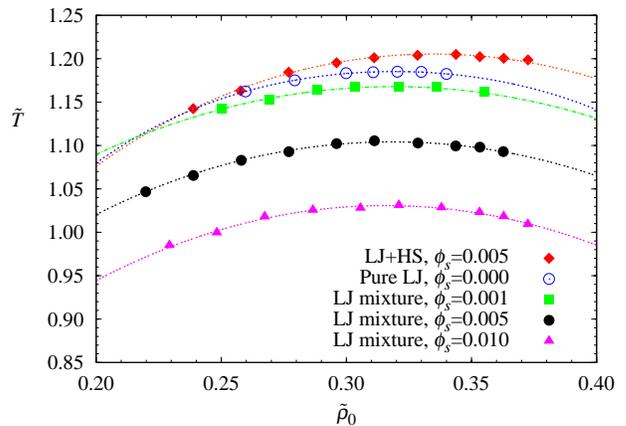}
  \caption{(color online) Iso-$Q^*$ curves for size-asymmetric binary
  mixtures consisting of a LJ fluid (particle size~$\sigma$) and
  small additives (particle diameter $\sigma/10$, volume
  fraction~$\phi_s$).  All curves pertain to a linear system size
  $L=8.06\sigma$. For additives that interact with the fluid as hard
  spheres (diamonds, upper curve), the critical temperature and density
  increase compared to the pure fluid (open circles), whereas the
  critical temperature decreases strongly if the additives have a weak
  attraction with the fluid (squares, filled circles, triangles).}
  \label{fig:mixtures}
\end{figure}

As before, we perform simulations for a range of $\rho_0$, at
\emph{fixed} additive volume fraction $\phi_s = \frac{\pi}{6} \sigma_s^3
\rho_s = 0.005$, corresponding to $N_s = 10000$. Because the small
particles are so numerous, and disperse relatively homogeneously, the
insertion probability of a fluid (large) particle in a standard GC
approach would be prohibitively small. By contrast, the present scheme
renders it feasible to equilibrate the system and sample the density
fluctuations.  Figure~\ref{fig:mixtures} (diamonds) shows that the
iso-$Q^*$ curve (plotted as a function of the \emph{total} reduced
density $\tilde\rho_{0} \equiv \rho_l \sigma^3 + \rho_s \sigma_s^3$) has
a parabolic shape, as for the pure fluid. However, despite the
relatively small volume fraction of additives, the maximum of this
curve, i.e., the liquid-vapor critical point of the mixture, is markedly
shifted. The increase in the critical temperature reflects an enhanced
attraction between the large particles which stems from the entropic
depletion interactions induced by the additives~\cite{asakura54}.

To highlight the subtle role of the nature of the \emph{interactions} of
the additives with the larger species, as well as their concentration,
we study several systems in which there is a weak attraction between
large and small particles. The interaction is again of the LJ form,
Eq.~(\ref{eq:lj-small}), in which $\sigma_s$ is replaced with
$\sigma_{ls}$.  Already at $\phi_s = 10^{-3}$ the critical temperature
is now noticeably \emph{depressed}, as is evident from the shift of the
iso-$Q^*$ maximum in Fig.~\ref{fig:mixtures}, and at $\phi_s = 10^{-2}$,
$T^c$ has decreased by almost 20\%! We explain this surprising effect by
the formation of a shell of small particles around the large particles,
akin to nanoparticle halo formation~\cite{tohver01,liu04b,martinez05},
which weakens the effective attraction between the large particles.

\begin{figure}
  \includegraphics*[width=\figurewidth]{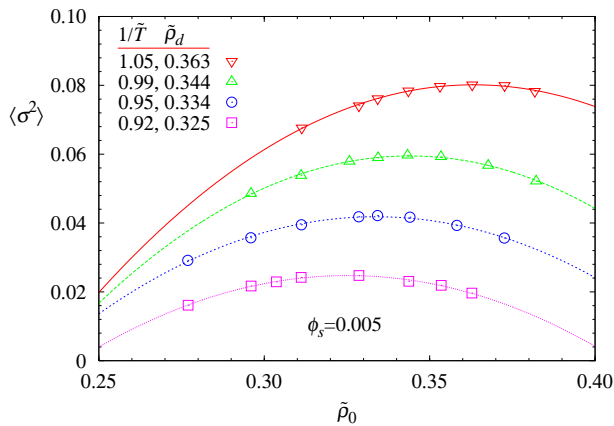}
  \caption{(color online) Variance of $P^{RG}(x)$ for a LJ mixture with
  size asymmetry $\sigma/\sigma_s=10$, as a function of total number
  density~$\tilde\rho_0$. Each (isothermal) curve is obtained from
  simulations at selected densities, but with constant additive volume
  fraction~$\phi_s=0.005$, and is parabolic (see fits), confirming
  Eq.~(\ref{eq:m2-parabola}). The maxima locate the coexistence curve
  diameter.}
  \label{fig:m2}
\end{figure}

\begin{figure}
  \includegraphics*[width=\narrowfigurewidth]{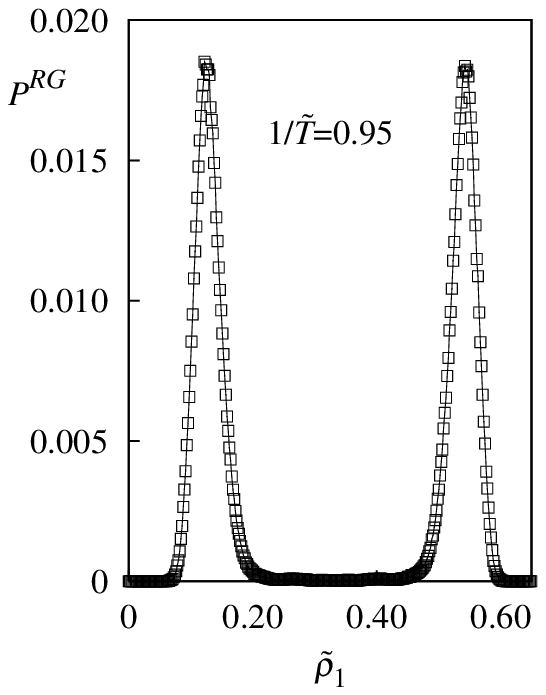}
  \includegraphics*[width=\narrowfigurewidth]{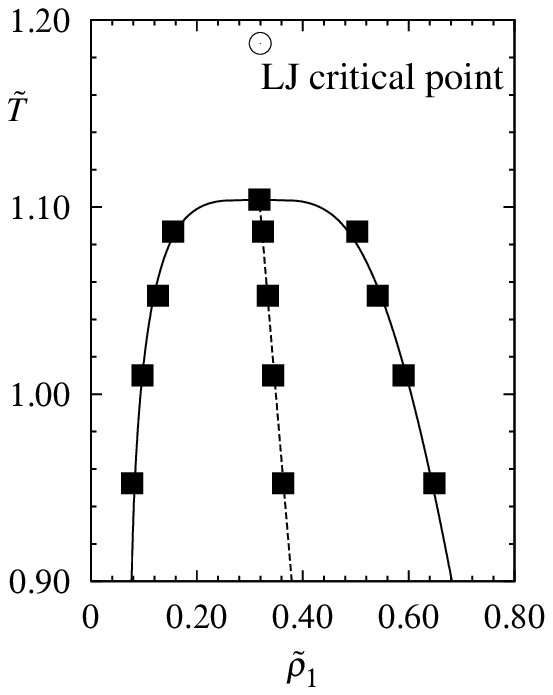}
  \caption{(a) Example of a $P^{RG}$ at $\rho_0=\rho_d$, for the fluid
  mixture described in Fig.~\ref{fig:m2}. (b) Corresponding phase
  diagram; the open circle indicates the critical point of a pure LJ
  fluid. The full line is a fit of the form $\rho^{\pm}-\rho^c =ut+v
  t^\beta$.}
  \label{fig:coex}
\end{figure}

Our approach not only yields accurate estimates of critical points, but
also entire coexistence curves. As described above, for each subcritical
temperature, the variance of $P^{RG}(x)$ has a maximum at the
coexistence curve diameter~$\rho_d$ [Eq.~(\ref{eq:m2-parabola})], as is
confirmed in Fig.~\ref{fig:m2}. The (total) densities of the coexisting
liquid and vapor phases are determined from the peak positions of
$P^{RG}$ [see Fig.~\ref{fig:coex}(a) for an example], resulting in the
phase diagram in Fig.~\ref{fig:coex}(b). We emphasize that obtaining
such a phase diagram in a reasonable timescale would not be feasible
using even the most efficient traditional approach to fluid phase
equilibria, namely GC simulation \cite{wilding95}. Our tests show that
the GC relaxation time is too large to be reliably estimated.
Nevertheless, a lower bound on the GC relaxation time, relative to that
of the pure LJ fluid, can be estimated via a comparison of the
large-particle transfer (insertion/deletion) acceptance probability
$p_{\rm acc}$. For liquid-like densities of the large particles
($\tilde\rho_1\approx 0.6$), we find that for $\phi_s=0.005$, $p_{\rm
acc}\sim 10^{-4}$; while for $\phi_s=0.01$, this falls to $p_{\rm
acc}\sim 10^{-6}$. These values are to be compared with $p_{\rm acc}\sim
10^{-1}$ for the pure LJ fluid. One can therefore expect the GC
relaxation time of the mixtures we have studied to be several orders of
magnitude greater than for the pure LJ fluid. Since the algorithm
presented here operates along fundamentally distinct lines---proposing
large-scale collective updates of clusters of small and large particles,
and accepting them with {\em unit} probability---it is not hampered by
this problem. Consequently it allows the efficient calculation of phase
diagrams, even under conditions for which the GC approach fails.

Summarizing, we have extended the rejection-free GCA to the study of
phase transitions, by embedding it within a restricted Gibbs ensemble.
The accurate location of critical points and coexistence curves within
this ensemble requires a suitable FSS theory, which has been presented
as well. By means of illustration, we have applied our method to a
strongly size-asymmetric LJ mixture, which cannot be studied with
existing direct methods, e.g., the GC ensemble.  We find that the
liquid-vapor phase behavior is highly sensitive to the concentration of
small particles and the nature of their interaction with the large ones.
Thus our method should prove useful in predicting the alterations to
phase behavior which occur when small particles of various types are
added to a fluid. Furthermore, by employing the method with a variant of
the GCA suitable for electrostatic
interactions~\cite{note-electrostatics}, it becomes possible to study
the effects of adding salt on the phase behavior of charged colloids.
Finally, we note that while for concreteness we have developed the
formalism for the case of phase transitions whose order parameter is the
density, the structure of our theory holds also for consolute points, or
indeed situations where the order parameter is a linear combination of
density and concentration.

\begin{acknowledgments}
  This material is based upon work supported by the National Science
  Foundation under Grant DMR-0346914 (to EL) and by EPSRC Grant
  GR/S59208/01 (to NBW).
\end{acknowledgments}


\begin{thebibliography}{10}

\bibitem{larson99}
R.~G. Larson, {\em The Structure and Rheology of Complex Fluids} (Oxford
  University Press, Oxford, U.K., 1999).

\bibitem{frenkel-smit2}
D. Frenkel and B. Smit, {\em Understanding Molecular Simulation}, 2nd ed.
  (Academic, San Diego, 2002).

\bibitem{tohver01}
V. Tohver {\it et~al.}, Proc. Natl. Acad. Sci. U.S.A. {\bf 98},  8950  (2001).

\bibitem{dress95}
C. Dress and W. Krauth, J. Phys. A {\bf 28},  L597  (1995).

\bibitem{geomc}
J. Liu and E. Luijten, Phys. Rev. Lett. {\bf 92},  035504  (2004).

\bibitem{liu05a}
J. Liu and E. Luijten, Phys. Rev. E {\bf 71},  066701  (2005).

\bibitem{liu04b}
J. Liu and E. Luijten, Phys. Rev. Lett. {\bf 93},  247802  (2004).

\bibitem{martinez05}
C.~J. Martinez {\it et~al.}, Langmuir {\bf 21},  9978  (2005).

\bibitem{panagiotopoulos87}
A.~Z. Panagiotopoulos, Mol. Phys. {\bf 61},  813  (1987).

\bibitem{mon92}
K.~K. Mon and K. Binder, J. Chem. Phys. {\bf 96},  6989  (1992).

\bibitem{bruce97}
A.~D. Bruce, Phys. Rev. E {\bf 55},  2315  (1997).

\bibitem{binder81}
K. Binder, Z. Phys. B {\bf 43},  119  (1981).

\bibitem{bruce92}
A.~D. Bruce and N.~B. Wilding, Phys. Rev. Lett. {\bf 68},  193  (1992).

\bibitem{note-longpaper}
J. Liu, N.~B. Wilding, and E. Luijten, to be published.

\bibitem{tsypin00}
M.~M. Tsypin and H.~W.~J. Bl{\"o}te, Phys. Rev. E {\bf 62},  73  (2000).

\bibitem{note-diameter}
This is confirmed by inserting $\rho_0=(\rho_g+\rho_l)/2$ into
  Eq.~({\ref{eq:map}}); clearly when $\rho_1=\rho_g$, one has
  $2\rho_0-\rho_g=\rho_l$ and thus both $P(\rho_1|t)$ and $P(2\rho_0-\rho_1|t)$
  are maximized, implying that so too is $P^{RG}(\rho_1)$. The same holds for
  $\rho_1=\rho_l$.

\bibitem{wilding95}
N.~B. Wilding, Phys. Rev. E {\bf 52},  602  (1995).

\bibitem{asakura54}
S. Asakura and F. Oosawa, J. Chem. Phys. {\bf 22},  1255  (1954).

\bibitem{note-electrostatics}
S.~A. Barr and E. Luijten, to be published.

\end{thebibliography}
\end{document}